\documentclass[a4paper,12pt]{article}
\usepackage[mathletters]{ucs}
\usepackage[utf8x]{inputenc}
\usepackage[fleqn]{amsmath}
\usepackage{amssymb,amsthm}
\usepackage[noend]{algorithmic}
\usepackage{algorithm}

\usepackage[verbose]{yogi_labels}
\usepackage[terse]{yogi_algorithm}
\usepackage{yogi_lang}
\usepackage{xth}
\newcommand{\ST}{\:|\:}
\newcommand{\Abs}[1]{\ensuremath{\left|#1\right|}}

\newcommand{\Set}[1]{{\{#1\}}}

\newcommand{\str}[1]{{\textit{\texttt{#1}}}}
\newcommand{\alg}[1]{{\textup{\textsf{#1}}}}
\newcommand{\Acronym}{\ensuremath{\mathcal{S}}}
\newcommand{\BWT}{\alg{BWT}}

\numberwithin{algorithm}{section}

\title{A Bijective String Sorting Transform}
\author{Joseph (Yossi) Gil\footnote{\str{yogi@cs.Technion.ac.IL}} \\
        Google, Inc.\footnote{Work done while being on sabbatical leave
			       from the Technion---Israel Institute of Technology.} \\
        Haifa, Israel
	\and
	David Allen Scott \footnote{\str{David\_a\_Scott@email.com}} \\
        El Paso, Texas\\
        USA
}
\begin{document}

\maketitle

\begin{abstract}
Given a string of characters, the Burrows-Wheeler Transform
  rearranges the characters in it so as to produce another string
  of the same length which
  is more amenable to compression techniques such as move to
  front, run-length encoding, and entropy encoders.
We present a variant of the transform
  which gives rise to similar or better compression value,
  but, unlike the original, the transform we present is
  \emph{bijective}, in that the inverse transformation
  exists for all strings.
Our experiments indicate that using our variant of the transform
  gives rise to better compression ratio than the original Burrows-Wheeler
  transform.
We also show that both the transform and its inverse can be computed
  in linear time and consuming linear storage.
\end{abstract}

\section{Introduction}
Informally, the famous \emph{Burrows-Wheeler Transform} (BWT)~\cite{Burrows:Wheeler:94}
  can be described as follows.
\begin{quote}
\sl Given a string~$α$ of length~$n$, generate the~$n$  cyclic rotations
  of~$α$, and then sort these.
By picking the last character of these sorted strings we obtain
  a string~$\BWT(α)$,
  the Burrows-Wheeler transform of~$α$.
\end{quote}
BWT has become very popular for text
   compression application, because of two important properties.
\emph{First}, if~$α$ is textual, then~$\BWT(α)$ tends to have long runs of identical characters,
   which makes~$\BWT(α)$ more amenable to compression techniques such as run-length-encoding
   and move-to-front~\cite{Bentley:Sleator:Tarjan:Wei:86}.
Consider for example an innocent phrase such as
  “\str{now is the time for the truly nice people to come to the party}'',
  in which there are no runs of consecutive identical characters.
In contrast, the Burrows-Wheeler transform of this phrase becomes
\begin{quote}
\begin{verbatim*}
oewyeeosreeeepi mhchlmhp tttnt puio yttcefn  ooati       rrolt
\end{verbatim*}
\end{quote}
in which 24 characters participate in eight such runs, the longest of which comprises seven repetitions of the space character.
To see how these seven spaces are brought together, note that  the  phrase
  has seven words beginning with~“\str{t}'', and hence seven
  of its rotations begin with “\str{t}''.
When sorted, these seven rotations become neighbors, and since the last character
   in these seven is space, seven such spaces occur consecutively in the transform.

The \emph{second} important property of the transform is the (rather surprising) fact that it is invertible
 in the following sense:
Given~$\BWT(α)$, it is possible to
  efficiently generate all rotations of~$α$.

The main issue with the inversion that concerns us here
  is that, by definition,~$\BWT(α)=\BWT(α')$
  for every~$α'$ which is a  rotation of~$α$.
Therefore, with the absence of additional information, regenerating~$α$ given~$\BWT(α)$
  is impossible.

To make decompression possible, compression algorithms must therefore store
   not only the compressed representation of~$\BWT(α)$, but also the \emph{rotation index}, an integer~$i$,~$0 ≤ i < n$,
   of~$α$ in the sorted list of its cyclic rotations.
A less “pure'' alternative is to append a newly introduced
   end-of-string character~$\square$ to~$α$, and then compute~$\BWT(α\square)$.
String~$α$ is then chosen as the rotation in which~$\square$ comes last.

This issue, together with a simple counting argument, shows that~$\BWT^{-1}(\cdot)$,
   the inverse Burrows-Wheeler transform, cannot be defined for all
   strings.
In fact, if~$η$ is  a string of length~$n$ selected at random,
   then~$\BWT^{-1}(η)$ is defined with probability~$1/n$.

A natural question is \emph{whether there exists a similar transform which
  is truly invertible}, in the sense that the transformed string uniquely identifies the original.
In this paper, we answer this question affirmatively  by describing a bijective,
  string sorting transform~$\Acronym(\cdot)$, which is similar to the Burrows-Wheeler transform
  in that it tends to bring identical characters together.
In fact, in many cases, the output is quite similar to that of the~$\BWT$ transform.
For example, applying~$\Acronym$ to the above phrase  yields
  a string which is different in only six locations:
\begin{quote}
\begin{verbatim*}
yoeyeeosreeeepi mhchlmhp tttnt puio wttcefn  ooati       rrotl
\end{verbatim*}
\end{quote}
Moreover, the~$\Acronym$ transform features in this case
  the same number of runs of identical characters and the same number
  of identical characters participating in these runs.
Our experimental results indicate that
  compression revolving around~$\Acronym$ tends to perform (slightly better) than~$\BWT$
  based compression.

Consider for example, \Tab{comparison}  which
compares the performance of the~$\Acronym$ transform  with that of the~$\BWT$
transform  when used for compression the famous Calgary corpus, a collection of 18 files which serves as the de-facto standard
  for benchmarking compression algorithms.
In preforming the measurements, we first used Nelson's reference implementation~\cite{Nelson:96}, which carries out
compression in five steps:
\emph{(i)}~initial run-length-encoding,
\emph{(ii)}~Burrow-Wheeler transformation
\emph{(iii)}~move-to-front
\emph{(iv)}~yet another run-length-encoding
\emph{(v)}~arithmetical encoding~\cite{Witten:Neal:Cleary:87}.
We then repeated the same steps, substituting~$\Acronym$-transform
 for~$\BWT$ in step~\emph{(ii)}.
(In this particular experiment, we relied on the bit- rather than byte- representation of the data.
The results for byte based compression are similar~\cite{Nagy:06}).

\begin{table}[hbt]
\begin{center}
\begin{tabular}{| l | r | r r| r r| r r|}
\hline
\textbf{File} &  \multicolumn{1}{|c|}{\textbf{Size}}
                  & \multicolumn{2}{|c|}{\textbf{\BWT-compression}}
                  &   \multicolumn{2}{|c|}{\textbf{\Acronym-compression}}
                  &  \multicolumn{2}{|c|}{\textbf{Gain}} \\
 & & \small{bytes}  & \small{ratio} &  \small{bytes} & \small{ratio}& \small{absolute} & \small{relative} \\
\hline
\textsc{BIB}   &   111,261 &   32,022  &   28.78\%  &   31,197  &   28.04\%  &   0.74\%  & 2.58\%   \\
\textsc{BOOK1}   &   768,771 &   242,857 &   31.59\%  &   235,913 &   30.69\%  &   0.90\%  &  2.86\%   \\
\textsc{BOOK2}   &   610,856 &   170,783 &   27.96\%  &   166,881 &   27.32\%  &   0.64\% &  2.28\%   \\
\textsc{GEO}   &   102,400 &   66,370  &   64.81\%  &   66,932  &   65.36\%  &   -0.55\% &  -0.85\%  \\
\textsc{NEWS}   &   377,109 &   135,444 &   35.92\%  &   131,944 &   34.99\%  &   0.93\% &  2.58\%   \\
\textsc{OBJ1}   &   21,504  &   12,727  &   59.18\%  &   12,640  &   58.78\%  &   0.40\% &  0.68\%   \\
\textsc{OBJ2}   &   246,814 &   98,395  &   39.87\%  &   94,565  &   38.31\%  &   1.55\%  &3.89\%   \\
\textsc{PAPER1}   &   53,161  &   19,816  &   37.28\%  &   18,931  &   35.61\%  &   1.66\%  & 4.47\%   \\
\textsc{PAPER2}   &   82,199  &   28,084  &   34.17\%  &   27,242  &   33.14\%  &   1.02\% &  3.00\%   \\
\textsc{PAPER3}   &   46,526  &   18,124  &   38.95\%  &   17,511  &   37.64\%  &   1.32\% &  3.38\%   \\
\textsc{PAPER4}   &   13,286  &   6,047   &   45.51\%  &   5,920   &   44.56\%  &   0.96\% &  2.10\%   \\
\textsc{PAPER5}   &   11,954  &   5,815   &   48.64\%  &   5,670   &   47.43\%  &   1.21\%  & 2.49\%   \\
\textsc{PAPER6}   &   38,105  &   14,786  &   38.80\%  &   14,282  &   37.48\%  &   1.32\%  & 3.41\%   \\
\textsc{PIC}   &   513,216 &   59,131  &   11.52\%  &   52,406  &   10.21\%  &   1.31\%   &11.37\%  \\
\textsc{PROGC}   &   39,611  &   15,320  &   38.68\%  &   14,774  &   37.30\%  &   1.38\% &  3.56\%   \\
\textsc{PROGL}   &   71,646  &   18,101  &   25.26\%  &   17,916  &   25.01\%  &   0.26\%&   1.02\%   \\
\textsc{PROGP}   &   49,379  &   13,336  &   27.01\%  &   13,010  &   26.35\%  &   0.66\% &  2.44\%   \\
\textsc{TRANS}   &   93,695  &   22,864  &   24.40\%  &   22,356  &   23.86\%  &   0.54\%  & 2.22\%   \\  \hline
\emph{Total}   &   3,251,493   &   980,022 &   30.14\%  &   950,090 &   29.22\%  &   0.92\%  & 3.05\%   \\  \hline
\emph{Median}   &   76,923  &    21,340   &   36.60\%  &   20,644  &   35.30\%  &   0.94\% &  2.58\%   \\  \hline

\end{tabular}
\end{center}
\caption{Performance of BWT-based compression~$\Acronym$-based compression of the Calgary corpus.}
\label{Table:comparison}
\end{table}

As can be seen in the table, using the~$\Acronym$-transform, improves compression for all files except for \textsc{Geo}.
The gain in compression ratio is about 1\%.
Note this gain is much greater than what can be attributed to the saving due
  due bijectivity, that is, the elimination of the end-of-string character, or
  the rotation index: Even in the \textsc{ProgL} file, in which the relative gain
  of the~$\Acronym$-based compression is the smallest, the size saving is
  of almost 200 bytes.

Other than better compression,~$\Acronym$ offers several other
  advantages over~$\BWT$.
First, there is no need to store an end-of-string marker, nor the rotation index,
  in applying the Burrows-Wheeler transform in loseless compression algorithms.
This advantage is prominent especially if the transform is used for very
  short texts, e.g., in transforming
  separately each line in a text file, or each field in database.
Second, since the algorithm is bijective, it is more adequate for application in which the compressed data
  is encrypted---a non-bijective transform necessarily reveals information to the attacker.
Finally, some may appreciate the elegance in bijectiveness and in the details of the definition of the transform.

For the impatient reader, an informal (and imprecise) description of the~$\Acronym$ transform
  is as follows:
\begin{quote}
\sl Break~$α$ into sub-strings by successively selecting and removing its “smallest'' suffix.
         Generate the rotations of each such sub-string, and sort all these rotations together.
        The transform~$\Acronym$ is then obtained by taking the “last'' character of this sorted list.
\end{quote}
Missing pieces in the above description include: the exact definition of the manner in which suffixes
  are compared, the specification of the order relation between rotations of different length,
  elaborating the meaning of the phrase “last'' character.
More importantly, we will also need to explain why this, seemingly arbitrary process,
  is reversible.

The transform~$\Acronym$ was discovered by the second author but remained unpublished.
Unfortunately, his announcements on Internet Usenet
  groups such as \str{comp.compression} on December 2007
  were received with great skepticism regarding issues including feasibility, correctness,
  complexity and utility.
Here we formalize and describe the algorithm in detail, prove its correctness,
  provide a linear time and space implementation, compare to related word,
  and discuss extensions ad generalizations.

\paragraph{Outline} The remainder of this article is structured as follows.
\Sec{BWT}gives some basic notations and demonstrates these in a precise
  definition of the algorithm for implementing Burrows-Wheeler transform.
Then, the reader is reminded of the
  linear time algorithm for inverting the transform.
We use the description of these two algorithms next,  in \Sec{BWTS} for defining the~$\Acronym$ transform,
  and in \Sec{IBWTS},  the algorithm for
  implementing its inverse.
In \Sec{Correctness} we explain why the algorithm for computing~$\Acronym^{-1}$
  is indeed correct.
\Sec{Final} concludes

\section{Preliminaries I: The Burrows-Wheeler Transform}
\label{Section:BWT}
This section serves as a reminder of the details of the Burrows-Wheeler
  transform.
It also sets up some definitions to be used later.
Let~$Σ$ be an ordered set, the \emph{alphabet} of \emph{characters},
  and let~$Σ⁺$ be the set
  of finite non-empty strings of characters chosen from~$Σ$.
For a  string~$α∈Σ$, let~$\Abs{α}$, the \emph{length} of~$α$,
  be the number of characters in~$α$.
The order relation in~$Σ$ is extended to a total order of
  strings of \emph{equal length} in~$Σ⁺$,
  in the usual lexicographical way.
At this stage, we leave the comparison of strings of non-equal length
  unspecified.

We will treat strings as arrays in the \C programming language, so~$α[0]$ shall
  denote the first
  character of~$α$,~$α[1]$  its second character, etc.
Further, let~$α[-1]$ denote the last character of~$α$,~$α[-2]$ its penultimate character,
  and more generally, for~$i ≥ \Abs{α}$ or~$i < 0$,  let~$α[i] ≡ α[i \bmod \Abs{α}]$.

For strings~$α, β∈Σ$, let~$αβ$ denote the string obtained by their concatenation; we say
  that~$α$ (respectively~$β>$) is a \emph{prefix} (respectively a \emph{suffix})
  of this concatenation.
For an integer~$m>0$ let~$α^m$ denote the string obtained by concatenating~$α$
  onto itself~$m$ times.

For~$α∈Σ⁺$ and an integer~$0 ≤ m ≤ \Abs{α}$
  let~$α(m)$ denote the~$m$\textsuperscript{\emph{th}} \emph{rotation}
  of~$α$, that is, the string obtained by removing the
  first~$m$ characters.of~$α$ and adding these at its end.
More precisely, if~$α=βγ$ and~$\Abs{α}=m$,
  then~$α(n) ≡ γβ$.
We extend this definition for all~$m ∈ \mathbb Z$
  by the equivalence~$α(m) ≡ α(m \bmod \Abs{α})$.
For example, if~$α=\str{tartar}$,
  then~$α(1)=α(4)=α(-2))=\str{artarta}$
  and~$α(2)=α(5)=α(-1)=\str{rtarta}$.
We have
 \begin{equation}
α(i)[j]=α[i + j]
\end{equation}
for all~$α ∈ Σ^+$ and~$i,j ∈ \mathcal Z$.

\Alg{Generate} describes, using these notations,
   the first step of the Burrows-Wheeler transform, that is,  the generation
   of the list of rotations of a given string.
\begin{algorithm}
\caption{$\alg{CyclicRotations}(α)$ \algorithmiccomment{Return the set of all cyclic rotations of~$α ∈ Σ⁺$.}}
\label{Algorithm:Generate}
\begin{algorithmic}[1]
\LET{n}{\Abs{α}}
\FOR{$i=0,…n-1$}
\STATE{$R←R ∪ \Set{α(i)}$}
\ENDFOR
\RETURN$R$
\end{algorithmic}
\end{algorithm}

\Alg{Generate} requires~$\mathcal O(n)$ time and storage:
Assuming that~$α$ is allocated in immutable
  storage, then each of~$α(i)$ can be represented by
  a triple of scalars: the address of the first character of~$α$,
  the index~$i$, and the length~$n=\Abs{α}$.
Henceforth, we  tacitly assume this triple based representation.

The second step of the~$\BWT$ transformation can now
  be described concisely as depicted by \Alg{Last}.

\begin{algorithm}[!hbt]
\caption{$\alg{Last}(R)$ \algorithmiccomment{\mbox{Given a set~$R ⊂ Σ^+$, return the string composed of the last character}}\newline
\mbox{}\hfill
 \algorithmiccomment{\mbox{of each of the members of~$R$, enumerated in lexicographical order.}}}

\label{Algorithm:Last}
\begin{algorithmic}[1]
\LET{n}{\Abs{R}}
\STATE{\textbf{let}~$η$ be an uninitialized string of length~$n$}
\FOR{$i=0,…, n-1$}
  \LET{α}{\min R}
  \STATE{$η[i]←α[-1]$}
  \STATE{$R←R ∖ \Set{α}$}
\ENDFOR
\RETURN$η$
\end{algorithmic}
\end{algorithm}

Note that the algorithm is tantamount to sorting the input set~$R$.
If~$R$ has~$n$ elements, then this sorting can be done in~$\mathcal O(n \log n)$
string comparisons.
Each such comparison may require~$\mathcal O(n)$  character comparisons, leading
  to an~$\mathcal O(n² \log n)$ implementation.
Yet, as Giancarlo, Restivo and Sciortino~\cite{Giancarlo:Restivo:Sciortino:07}
  observe, in the case that~$R$ is indeed a set of rotations, then the
  sorting can be done in~$\mathcal O(n)$ time, by reduction to the problem of sorting
  the suffixes of a given string, which is known to be linear time.

Functions~$\alg{CyclicRotations}$ (\Alg{Generate}) and~$\alg{Last}$  (\Alg{Last}) are combined
  in \Alg{BW}, which realizes the Burrows-Wheeler transform.
Evidently, the algorithm requires linear time and space.

\begin{algorithm}[!hbt]
\caption{$\BWT(α)$\algorithmiccomment{Given a string~$α ∈ Σ⁺$, return its Burrows-Wheeler transform.}}
\label{Algorithm:BW}
\begin{algorithmic}[1]
\STATE{$R←\alg{CyclicRotations}(α)$}
\RETURN$\alg{Last}(R)$
\end{algorithmic}
\end{algorithm}

\section{Preliminaries II: Inverting the Burrows-Wheeler Transform}
\label{Section:IBWT}
Let~$α∈ Σ⁺$ be a string of length~$n$,
  and let~$η=\BWT(α)$, then, examining \Alg{BW}, we see that it
  effectively defines a permutation~$π$,
  such that~$η[i]=α[π(i)]$ for~$i=0,1,…, n-1$.
Given~$η$, we would like to generate the inverse permutation~$π^{-1}$.
Unfortunately, as explained above,  this is impossible.

Instead, the algorithm conceived by Burrows and Wheeler
  produces from~$η$ a permutation~$θ$ from
  which a \emph{rotation} of~$α$ can be generated.
The defining property of permutation~$θ$ is
\begin{Equation}[theta]
  ∀ k \bullet (0 ≤ k < n) \wedge  (k=π(i)) \; \Longrightarrow  \; θ(k)=π(i-1 \bmod n).
\end{Equation}%
That is, having matched a position~$i$ in~$α$
  with a position~$k$ in~$η$,
    we can match,~$i -1 \bmod n$, the cyclically preceding location in~$α$
  with the position~$θ(k)$ in~$η$.
Let us therefore define a permutation~$ρ$ by applying~$θ$
  upon itself successively, \emph{i.e.}, \[
    ρ(i)=\begin{cases}
                       θ(0) & i=0 \\
                       θ(ρ(i-1)) & i > 0.
                      \end{cases}.
 \]

Applying~$ρ$ to reorder the characters  in~$η$ generates,
  last to first, the characters of some rotation of~$α$.
More precisely, Burrows and Wheeler's inversion procedure generates the string~$β$,
  defined by~$β[j]=η[ρ(n-j)]$,
  which is not only \emph{a} cyclic rotation of~$α$, it is the  \emph{lexicographically smallest} such
  rotation.

The full process of generating~$β$ from~$θ$ is described in \Alg{Thread}.

\begin{algorithm}[!htb]
\caption{$\alg{Thread}(η, θ)$
  \algorithmiccomment{
Given the transform~$η=\BWT(α)$, and the cyclic permutation~$θ$,}
  \newline
\mbox{}\hfill
  \algorithmiccomment{return  the string~$β$, the lexicographically smallest rotation of~$α$.}}
\label{Algorithm:Thread}
\begin{algorithmic}[1]
\LET{n}{\Abs{η}}
\STATE{\textbf{let}~$β$ be an uninitialized string of length~$n$}
\STATE{$k←0$}
\label{line:start}
\FOR[fill in~$β$,  last to first]{$i=n-1, n-2,…, 0$}
    \STATE{$β[i]←η[k]$}
\label{line:last}
    \STATE{$k←θ(k)$}
\ENDFOR
\RETURN$β$
\end{algorithmic}
\end{algorithm}

The algorithm is rather straightforward,
  except for Line~\ref{line:start}, which initiates the threading process
  from the first character for~$η$.
By doing that, we ensure that the smallest rotation of~$α$ is returned.
To see that, observe that the last character of this smallest rotation is the one mapped by the transform to~$η[0]$.
In mapping back this character to the last character of the output, as done in the first time Line~\ref{line:last}
  is executed, we ensure that the we generate precisely this rotation.

Concentrate on the sorted list of the rotations~$α(i)$,
  which we will denote by~$L$.
The following two lemmas establish the means for generating the permutation~$θ$ from~$η$.
\begin{Lemma}[next]
If for some~$i$,~$0 ≤ i <n$, rotation~$α(i+1)$ occurs at position~$k$ in~$L$,
  while~$α(i)$ occurs at position~$j$ in it, then~$j=θ(k)$.
\end{Lemma}

\begin{proof}
We have that~$η[k]=α(i+1)[-1]=α[i]$, and~$η[j]=α(i)[-1]=α[i-1]$.
Thus, if we knew that~$η[k]$ is mapped to a certain position in~$β$, we will be able to conclude
  that~$η[j]$ is mapped
  to the cyclically previous position in~$β$.
\end{proof}

Observe that \Lem{next} does not require the knowledge of~$i$.
All we need to know is that the rotation at position~$k$ is obtained by omitting the first
  character of the rotation at position~$j$, what is called by Burrows and Wheeler
  a \emph{match} between~$j$ and~$k$.

\begin{Lemma}[match]
For an arbitrary character~$c ∈ Σ$, consider
  the sorted list~$α(i₀),…, α(i_{ℓ-1})$ of those
  rotations~$α(i)$,~$0 ≤ i < n$,
  for which~$α(i)[-1]=c$ (that is, the rotations which correspond to occurrences of~$c$ in~$η$).
Then, the list~$α(i₀-1),…, α(i_{ℓ-1}-1)$  is also sorted.
  Moreover, this list occurs consecutively in~$L$.
\end{Lemma}

\begin{proof}
Since the first character of each of the rotations~$α(i₀-1),…, α(i_{ℓ-1}-1)$ is~$c$,
  we can rewrite these
  as~$c α(i₀),…, cα(i_{ℓ-1})$. This list is sorted since we assumed
 that~$α(i₀),…, α(i_{ℓ-1})$ are sorted.
Further, the elements of this list occur consecutively in~$L$ since they all begin with~$c$ and no other rotation
  begins with~$c$..
\end{proof}

\Lem{match} provides the means for matching the location of a rotation~$α(i)$ with the location of the
  rotation~$α(i-1)$.
To understand the process, consider first the case
  that~$c$ is the smallest character occurring in~$η$,
  and that it is found at locations~${k₀},…,{k_{ℓ-1}}$ in it.
We know that there are some~$i₀,…, i_{ℓ-1}$, such
  that~$α(i₀)[-1]=α(i₁)[-1]=⋯=α(i_{ℓ-1})[-1]=c$.
Also, the rotations~$α(i₀),…, α(i_{ℓ-1})$
  are sorted into into  locations~${k₀},…,{k_{ℓ-1}}$.
Although we do know the values~${i₀},…, i_{ℓ-1}$,  we can use \Lem{match} to
  infer the locations of the “preceding'' rotations~$α(i₀-1)…, α(i_ℓ-1)$:
By this lemma, these must occur in~$L$ together and at the same order.
Since~$c$ is the smallest character in~$η$, we can infer that these preceding rotations
  occur precisely at locations~$0, 1,…,ℓ-1$.
Matching the location in~$L$ of each~$α({i_j})$ with that of~$α(i_j - 1)$, and
  applying  \Lem{next}~$ℓ$ times we conclude
  that
\begin{Equation}[first]
        θ(k₀)=0,  θ(k₁)=1,…, θ(k_{ℓ-1})=ℓ-1.
\end{Equation}

Having done that, we can continue to the second smallest character occurring in~$η$,
  and repeat the process, except that this time, the
  preceding rotations must occur at
  location~$ℓ$ in~$L$.
So, if this character is found in locations~${k'₀},…,{k'_{ℓ'-1}}$, we have
\begin{Equation}[second]
        θ(k'₀)=ℓ, θ(k'₁)=ℓ+ 1,…, θ(k'_{ℓ-1})=ℓ+ℓ'-1.
\end{Equation}

\Alg{Match} applies this process to create~$θ$.
The algorithm uses characters of the alphabet~$Σ$ as array indices,
  tacitly assuming that~$Σ=\Set{0,…, \Abs{Σ}-1}$.

\begin{algorithm}
\caption{$\alg{Match}(η)$ \algorithmiccomment{
Given a string~$η ∈ Σ⁺$, return  the  permutation~$θ$.}}
\label{Algorithm:Match}
\begin{algorithmic}[1]
\LET{n}{\Abs{η}}
\COMMENT{determine the input's length}
\label{line:begin:mundane}
\STATE {\textbf{let} \alg{counts} be a zero initialized array of size~$\Abs{Σ}$}
\FOR[set~$\alg{counts}{[c]}=\Abs{\Set{i \ST η[i]=c}}$ for all~$c ∈  Σ$]{$i=0,…, n -1$}
    \LET{c}{η[i]}\COMMENT{the character in the input we currently inspect}
    \STATE{$\alg{counts}[c]←\alg{counts}[c]+ 1$}
    \COMMENT{count this occurrence of~$c$}
\ENDFOR
\STATE {\textbf{let} \alg{before} be a zero initialized array of size~$\Abs{Σ}$}
\FOR[set~$\alg{before}{[c]}=\Abs{\Set{i \ST η[i] < c}}$  for all~$c ∈  Σ$]{$c=2,…, \Abs{Σ}$}
    \STATE~$\alg{before}[c]=\alg{before}[c-1]  +  \alg{before}[c]$
    \COMMENT{standard prefix sum}
\ENDFOR

\label{line:end:mundane}
\STATE {\textbf{let} \alg{seen} be a zero initialized array of size~$\Abs{Σ}$}
\FOR[set~$θ(i)$ to the next available match]{$i=0,…, n -1$}
\label{line:begin:heart}
  \LET{c}{η[i]}\COMMENT{the character in the input we currently inspect}
  \STATE{$θ(i)←\alg{before}[c] + \alg{seen}[c]$}
  \COMMENT{locations~$0,…,\alg{before}[c]-1$ are reserved for~$c'<c$,}\newline
\label{line:heart}
  \COMMENT{while locations~$\alg{before}[c],…, \alg{before}[c] + \alg{seen}[c] -1$}\newline
  \mbox{}\COMMENT{were used for earlier occurrences of~$c$}
  \STATE{$\alg{seen}[c]←\alg{seen}[c] + 1$}
  \COMMENT{mark this occurrence of~$c$ as seen}
\ENDFOR
\label{line:end:heart}
\RETURN{$θ$}
\end{algorithmic}
\end{algorithm}

Lines~\ref{line:begin:mundane} through~\ref{line:end:mundane}
  in the algorithm are mundane; their main purpose is to compute the contents of
  array~$\alg{before}$, which, at its~$c$\xth position contains the number
  of times a character strictly  smaller than~$c$ occurs in the in the input.

The heart of the algorithm is in lines~\ref{line:begin:heart} through~\ref{line:end:heart}.
This loop effectively implements the process described above for each
  of the characters that occur in~$η$.
The tricky part is that this is done simultaneously for all characters.
Thus, instead of iterating over the different characters in~$η$, and
  then, examining for each character all its locations, the loop in
  line~\ref{line:begin:heart} scans the positions in~$η$ in order.
Array~$\alg{seen}$ records at position~$c$, the number of
  times that~$c$ was seen in course of this scan.

Line~\ref{line:heart} is the essence of the loop;
  this line generalizes~\eq{first} and~\eq{second}.
The value of \alg{before[c]} provides the baseline, that is, the locations which
  are reserved for smaller characters
  (these locations were matched in previous iterations, or will be matched
  by subsequent iterations of this loop), while \alg{seen[c]}
   is the number of matches of~$c$-locations
 which were recorded in previous iterations of this loop into~$θ$.

Finally, \Alg{UNBWT}, combines functions \alg{Thread} (\Alg {Thread}) and \alg{Match} (\Alg{Match})
  for inverting the Burrows-Wheeler transform.

\begin{algorithm}
\caption{$\BWT^{-1}(η)$
   \algorithmiccomment{\mbox{For a string~$η ∈ Σ⁺$, return  the smallest string~$β$,such that~$\BWT(β)=η$}}}
\label{Algorithm:UNBWT}
\begin{algorithmic}[1]
\LET{θ}{\alg{Match}(η)}
\RETURN$\alg{Thread}(η,θ)$
\end{algorithmic}
\end{algorithm}

\section{The Bijective String Sorting Transform}
\label{Section:BWTS}
In this section, we present the~$\Acronym$-transform, our bijective- Burrows-Wheeler- like
  string sorting transform and its inverse~$\Acronym^{-1}$.
The outline of the algorithm for computing~$\Acronym$  is similar to that of the Burrows-Wheeler including computing rotations, sorting these,
  and then selection of the last character.

The main difference is that the~$\Acronym$-transform  does not work on the entire input
  as a whole.
Instead, given a string~$α$, the transform decomposes it into words,
\begin{Equation}[decompose]
    α=ω₀ ω₁ ⋯ ω_{m-1}
\end{Equation}%
and then proceeds to computing
   the rotations of each of these words, sorting all of the rotations
    together, and then selecting the last character of the rotations in their sorted order.
The details are supplied  \Alg{Acronym}.

\begin{algorithm}[!htp]
\caption{$\Acronym(α)$\algorithmiccomment{Given a string~$α∈Σ⁺$, return the
  bijective string sorting transform~$\Acronym(α)$.}}
\label{Algorithm:Acronym}
\begin{algorithmic}[1]
\STATE{$W←\alg{Factor}(α)$}\COMMENT{compute the Lyndon factorization of~$α$}
\label{Line:decompose}
\STATE{$R←∅$} \COMMENT{$R$ will be the set of rotations of these fragments}
\FORALL[retrieve all rotations of~$ω$]{$ω ∈ W$}
\STATE{$R←R ∪ \alg{CyclicRotations}(ω)$}
\COMMENT{and collect these into~$R$}
\ENDFOR
\RETURN~$\alg{Last}(R)$
\label{Line:last}
\end{algorithmic}
\end{algorithm}

The algorithm uses as subroutines function
  \alg{CyclicRotations} (presented above in \Alg{Generate})
  to produce all the cyclic rotations of the fragments,
  and function \alg{Last} (presented above in \Alg{Last})
  for sorting these, and selecting their last element.

The factorization \eq{decompose}
 is such that each~$ω_i$ is a Lyndon~\cite{Lyndon:54},
  i.e., a word which  is smaller than all of its rotations,~$ω_i < ω_i(j)~$ for~$j=1,2,…, \Abs{ω_i}-1$.
It is also required that~$ω₀ < ω₁ <  ⋯ < ω_{m-1}$.
The  presentation of~$α$ in the
  form~\eq{decompose} ssatisfying these two properties is known as the \emph{Lyndon factorization}.
It is well known that the Lyndon factorization  is unique.
Function \alg{Factor} called in~\ref{Line:decompose} in the \Alg{Acronym} uses
  Duval's~\cite{Duval:83}.algorithm for computing the Lyndon factorization in linear time and space.

Recall that we have left open the issue of extending the order relation in~$Σ$ to
  strings of unequal length in~$Σ^+$.
Our transform works with two possible such extensions, the usual lexicographical    comparison
  in which if~$α$ is a prefix of~$β$ then~$α < β$.
The other extension, which can be viewed as slightly more elegant, is
  that in comparing two string of unequal length,
  we compare the \emph{infinite} periodic repetitions of  each of these,
  or, phrased differently,
  comparing~$α^{\Abs{β}}$ with~$β^{\Abs{α}}$.
Consider for example the strings  “\str{the}''' and “\str{there}''.
Then,  \(\str{the} < \str{there} \) according to the first definition,
  while \(\str{the} > \str{there} \)  according to the \emph{infinite-periodic}
  order
  since \(\str{the}⋯ > \str{therethere}⋯\).

Interestingly, the Lyndon factorization algorithm works for both
  variations of the “lexicographical'' orders.
The point where these differ is in sorting together the
  the rotations of these Lyndon words.
For the first, simple and standard (but somewhat less elegant), definition
  we can sort the rotations together using the linear time suffix array construction algorithm of Kärkkäinen, Sanders and
  Burkhardt~\cite{Karkkainen:Sanders:Burkhardt:06}.

Recall that Kärkkäinen et.\ al's  algorithm sorts the suffixes recursively, where in
  each recursive step, the algorithm partitions the input into character triples,
  where each character triple is considered a new character.
A linear time, radix sort, algorithm is applied to the new set of characters,
  and the recursion continues only if these new characters are not all distinct.
In applying this algorithm to sorting the rotations of a single, non-periodic
  string (as we have in the Lyndon decompositions)~$ω$,
  it is sufficient to sort the suffixes of~$ωω$.
Thus, we can sort set of rotations of a single Lyndon factor
  separately.

Consider now the problem of comparing rotations of
  of the factors~$ω₀, ω₁,…, ω_{m-1}$, where we
  leave aside the issue of comparing rotations of the same Lyndon factor.
We deal with this specific problem using the same paradigm
  of  Kärkkäinen et.\ al's, except that the grouping together of the
  triples is a cyclic fashion is done in a cyclic fashion.
Of course, if the size of some~$ω_i$ is not divisible by three,
  the recursive step does \emph{not} reduce the number
  of characters in it.
In the~$j$\textsuperscript{\emph{th}} recursive step, the algorithm thus
  manipulates “characters'' which belong to sequences of length~$3^j$
  in the input.
To ensure that only linear work is done, we prune at the~$j$\textsuperscript{\emph{th}}
  step those characters which belong to factors whose length is no greater than~$3^j$.
This pruning is carried out \emph{even} if these characters are not unique!

Merging the result of the inter-factor and intra-factor sorting steps
  can be easily done in linear time.
Unfortunately, this technique only works in linear time
   for the standard lexicographical
  order.
Sorting according to the infinite periodic order shall require~$O(n \lg n)$
  time.

\paragraph{Note.}\sl The Lyndon factorization probably accounts for the better compression
  results achieved by the~$\Acronym$-transform.
Recall that the standard application of BWT breaks the input into blocks, and then applies the “block-sorting'' procedure to each block.
In using~$\Acronym$-transform, we break each block into smaller blocks, the Lyndon words, and apply a similar (but not identical) process
to each such word.
Now, the fact that this refining breakdown into blocks is not arbitrary, but rather depends on underlying properties of the input,
  may very well be the reason for the better performance we witness.
\vspace{5pt}\rm

\section{Inverting the String-Sorting Bijective Transform}
\label{Section:IBWTS}
\Alg{Inverse:Acronym} gives
  the procedure for inverting the transform,~$\Acronym^{-1}$.

\begin{algorithm}[!hbt]
\caption{$\Acronym^{-1}(η)$
   \algorithmiccomment{For a string~$η ∈ Σ⁺$, return the string~$α$, such that~$\Acronym(α)=η$.}}
\label{Algorithm:Inverse:Acronym}
\begin{algorithmic}[1]
\LET{θ}{\alg{Match}(η)}
\RETURN$\alg{MultiThread}(η,θ)$
\end{algorithmic}
\end{algorithm}

Evidently, just like the inversion of the
  Burrows-Wheeler transform, the inversion~$\Acronym^{-1}$
  relies on an auxiliary permutation~$θ$, which plays a similar role
  in both inversions.
Rather surprisingly, the same function~$\alg{Match}$
  (recall \Alg{Match} above)
  can be used for generating the permutation~$θ$.
The difference is that this time~$θ$ is not cyclic.
Instead, applying~$\alg{Match}$ returns a permutation~$θ$ which
  has~$m$ cycles, each corresponding to a word~$ω_i$.

More specifically, traversing the cycle~$0, θ(0), θ(θ(0)),…, θ^{-1}(0)$ produces the
  word~$ω_{m-1}$, last to first character.
That is to say, if~$η=\Acronym(α)$,
  then the last character of~$w_{m-1}$ (and hence of~$α$)
  is~$η[0]$, the once preceding it is~$η[θ(0)]$, etc.

Let~$k$ now be the smallest integer which is not included in this first cycle.
Then, traversing the cycle~$k, θ(k), θ(θ(k)),…,  θ^{-1}(k)$
  produces~$ω_{m-2}$, again, from last character to first.
The remaining words are produced by carrying out this process iteratively.

\Alg{Inverse:Acronym} thus cannot use
  function \alg{Thread} (\Alg{Thread} above)
  to reconstruct the original string~$α$.
Instead, it uses a more general function,
  \alg{MultiThread},
  for traversing the permutation~$θ$.
Curiously,  function \alg{MultiThread} is  a true generalization of \alg{Thread},
  in the sense that the call to \alg{Thread} in the inverse Burrows-Wheeler transform,
 \Alg{UNBWT}, can be transparently be replaced by a call to \alg{MultiThread}.

The details of \alg{MultiThread} are depicted in \Alg{MultiThread}, which given
  a string~$η ∈ σ^+$,~$η=\alg{Acronym}(α)$, and a permutation~$θ$,
  with the properties as described above, traces the cycles in~$θ$, to produce the
  inverse transform~$α$, last character to first, out of~$η$.

\begin{algorithm}[!htb]
\caption{$\alg{MultiThread}(η, θ)$
  \algorithmiccomment{Given a string~$η=\Acronym(α)$, and
  the permutation~$θ$, return~$α$.
}}
\label{Algorithm:MultiThread}
\begin{algorithmic}[1]
\LET{n}{\Abs{η}}
\COMMENT{determine the input's length}
\label{line:multi:begin:mundane}
\STATE{\textbf{let}~$T$ be an uninitialized integers array of length~$n$}
\COMMENT{used for tracking cycles in~$θ$}
\FOR[initialize~$T$ with the permutation~$θ$]{$i=0,…, n-1$}
  \STATE{$T[i]←θ(i)$} \COMMENT{initialize the $i$\textsuperscript{\emph{th}} position}
\ENDFOR
\label{line:multi:end:mundane}

\STATE{\textbf{let}~$α$ be an uninitialized string of length~$n$}
\COMMENT{to be filled with the result, last  to first}
\STATE{$i←n-1$}
\COMMENT{$i$ is the next position in~$α$ to be filled}
\FOR[follow all cycles defined by~$θ$]{$j=0,…, n$}
\label{Line:outer:begin}
  \IF[a new cycle, starting at~$i$, was discovered]{$T[i] ≠ ⊥$}
    \STATE{$k←j$}  \COMMENT{$k$ is used for traversing the cycle beginning at~$i$}
    \REPEAT[traverse each element in the cycle which begins at~$i$]
\label{Line:inner:begin}
    \STATE{$α[i]←n[k]$} \COMMENT{produce the next output character}
    \STATE{$i←i-1$} \COMMENT{and step back to the next output character to fill}
    \STATE{$t←k$} \COMMENT{$t$ stores the previous value of~$k$}
    \STATE{$k←T[k]$} \COMMENT{proceed to the next element in the cycle}
    \STATE{$T[t]←⊥$} \COMMENT{mark previous element as visited}
    \UNTIL{$T[k]=⊥$} \COMMENT{the current cycle was exhausted}
\label{Line:inner:end}
  \ENDIF
\ENDFOR
\label{Line:outer:end}
\RETURN$α$
\end{algorithmic}

\end{algorithm}

Lines~\ref{line:multi:begin:mundane} through~\ref{line:multi:end:mundane} in this algorithm
  are mundane.
They produce a temporary array~$T$, which initially reflects the permutation~$θ$.
As we traverse the cycles of~$θ$, we mark each traversed element by setting the corresponding
  value of array~$T$ to~$⊥$.

Next, the algorithm proceeds to producing the returned string~$α$,
  starting at its last character, working its way to its first.
The outer loop (lines~\ref{Line:outer:begin}--\ref{Line:outer:end})
  examines each position in array~$T$ in turn.
If the value stored in this position is not~$⊥$, then the inner loop
  (lines~\ref{Line:inner:begin}--\ref{Line:inner:end})
  follows up the cycle in~$θ$ that starts at this position,
  producing an output character in each iteration, and marking each
  visited position by setting the corresponding location of~$T$ to~$⊥$.

\section{Correctness of the Inversion Algorithm}
\label{Section:Correctness}
It is easy to check that algorithms \Alg{Inverse:Acronym} and  \Alg{MultiThread}
  require linear time and space.
We now turn to the issue of their correctness.

Examining \Alg{Acronym}, we see that it effectively computes a
  permutation~$π$ of the input.
A position~$i$ in the input string~$α$ is first associated with
  a certain word~$ω ∈ W$.
Exactly one of the rotations of~$ω$ is such that this position comes to be the last
  character.
The sorting together of all rotations of the words in~$W$
  assigns an ordinal number to this rotation; this ordinal number is nothing else than~$π(i)$.

Given~$η=\Acronym(α)$, finding the inverse of~$π$
  is done again by computing the auxiliary permutation~$θ$,
  but this time,~$θ$ is defined in a piecemeal fashion.
Let~$π_ω: \Set{0,…, \Abs{ω}-1}→\Set{0,…, n-1}$ be the function
  describing the mapping
  from the positions of a word~$ω ∈ W$ into positions of~$η$
  as carried out by~$\Acronym$.
The defining property of~$θ$ is
\begin{Equation}[S]
  ∀ k, (0 ≤ k < n) \wedge  \left( k=π_ω(i) \right) \; \Longrightarrow  \; θ(k)=π(i-1 \bmod \Abs{ω}).
\end{Equation}\%
That is, having matched  a position~$k$ in~$η$ not only with some word~$ω$
  but also with a position~$i$ in that word,
    we can match position~$i-1 \bmod \Abs{ω}$, the cyclically preceding position, in~$ω$,
   with~$θ(k)$.

To understand why \Alg{Match} computes~$θ$ also for the~$\Acronym$ transform,
  let us consider the general setting in which we sort together rotations of multiple
  words.
Henceforth, let~$W⊂ {Σ}⁺$ be a fixed finite set of words,
  and let~$n=∑_{ω∈W}\Abs{ω}$, be the total
  length of all the words in~$W$.
Also, let~$L=L₀ L₁⋯ L_{n-1}$ be the sorted list of all rotations of the words in~$W$,
  so each~$L_i$ is a rotation of some word  in~$W$, and let~$η$ be the string
  defined by~$η[k]=L_k[-1]$.

The following generalizes \Lem{next}.
\begin{Lemma}[next:m]
If~$L_j=L_k(-1)$ then~$j=θ(k)$.
\end{Lemma}

\begin{proof}
The proof is essentially the same as that of the proof of \Lem{next}.
From the assumptions it follows that there is a word~$ω ∈ W$ and an index~$i$ such
  that~$L_k=ω(i+1)$ and~$L_j=ω(i)$.
The last character in~$L_k$ is therefore~$ω[i]$ while the last character in~$L_j$
  is~$ω[i-1]$.
Thus, if we knew that~$η[k]$ is mapped to a certain position in~$ω$, we will be able to conclude
  that~$η[j]$ is mapped
  to the cyclically previous position in~$ω$.
\end{proof}

\begin{Lemma}[match:m]
For an arbitrary~$c ∈ Σ$, let~$L_{i₀},…, L_{i_{ℓ-1}}$ be the list of
  all rotations~$L$ in  which~$c$ occurs as the last character, that is~$η[i₀]=η[i₁]=⋯=η[i_{ℓ-1}]=c$.
Then, the list~$L_{i₀}(-1),…, L_{i_{ℓ-1}}(-1)$ occurs consecutively and in that order in~$L$.
\end{Lemma}

\begin{proof}
Since the first character of each of the rotations~$L_{i₀}(-1),…, L_{i_{ℓ-1}}(-1)$ is~$c$,
  we can rewrite these
  as~$c L_{i₀},…, cL_{i_{ℓ-1}}$.
This list is sorted since we assumed
 that~$α(){i₀},…, τ^{i_{ℓ-1}}$ are sorted.
Further, the elements of this list occur consecutively in~$L$ since they all begin with~$c$ and no other rotation
  begins with~$c$..
\end{proof}

\section{Future Work}
\label{Section:Final}
Clearly, the work ahead of us is in evaluating the efficacy of the transform described here in
state of the art compression programs such as \emph{bzip2} and \emph{7-zip}.

We are intrigued by the question of sorting the rotations of the Lyndon decomposition
  in linear time with the \emph{infinite periodic} order.

\bibliographystyle{yogi_abbrv}
\bibliography{00,publishers,other_shorthands,yogi-book}

\end{document}